\documentclass[structabstract, longauth, regular]{aa}  
\usepackage{multicol}

\renewcommand{\epsilon}{\varepsilon}

\usepackage{graphicx}
\usepackage{txfonts}

\usepackage[switch]{lineno}
\usepackage{setspace}
\usepackage{hyperref}
\usepackage{caption}

\usepackage{natbib}
\bibpunct{(}{)}{;}{a}{}{,} 

\begin{document}
\title{Search for VHE gamma-ray emission from Geminga pulsar and
nebula with the MAGIC telescopes}

%

\author{
M.~L.~Ahnen\inst{1} \and
S.~Ansoldi\inst{2} \and
L.~A.~Antonelli\inst{3} \and
P.~Antoranz\inst{4} \and
A.~Babic\inst{5} \and
B.~Banerjee\inst{6} \and
P.~Bangale\inst{7} \and
U.~Barres de Almeida\inst{7,}\inst{24} \and
J.~A.~Barrio\inst{8} \and
J.~Becerra Gonz\'alez\inst{9,}\inst{25}
% ...
\and
}
\institute { ETH Zurich, CH-8093 Zurich, Switzerland
\and Universit\`a di Udine, and INFN Trieste, I-33100 Udine, Italy
\and INAF National Institute for Astrophysics, I-00136 Rome, Italy
\and Universit\`a  di Siena, and INFN Pisa, I-53100 Siena, Italy
\and Croatian MAGIC Consortium, Rudjer Boskovic Institute, University of Rijeka, University of Split and University of Zagreb, Croatia
\and Saha Institute of Nuclear Physics, 1/AF Bidhannagar, Salt Lake, Sector-1, Kolkata 700064, India
\and Max-Planck-Institut f\"ur Physik, D-80805 M\"unchen, Germany
\and Universidad Complutense, E-28040 Madrid, Spain
\and Inst. de Astrof\'isica de Canarias, E-38200 La Laguna, Tenerife, Spain; Universidad de La Laguna, Dpto. Astrof\'isica, E-38206 La Laguna, Tenerife, Spain
\and University of \L\'od\'z, PL-90236 Lodz, Poland
\and Deutsches Elektronen-Synchrotron (DESY), D-15738 Zeuthen, Germany
\and Institut de Fisica d'Altes Energies (IFAE), The Barcelona Institute of Science and Technology, Campus UAB, 08193 Bellaterra (Barcelona), Spain
\and Universit\"at W\"urzburg, D-97074 W\"urzburg, Germany
\and Universit\`a di Padova and INFN, I-35131 Padova, Italy
\and Institute for Space Sciences (CSIC/IEEC), E-08193 Barcelona, Spain
\and Technische Universit\"at Dortmund, D-44221 Dortmund, Germany
\and Finnish MAGIC Consortium, Tuorla Observatory, University of Turku and Astronomy Division, University of Oulu, Finland
\and Unitat de F\'isica de les Radiacions, Departament de F\'isica, and CERES-IEEC, Universitat Aut\`onoma de Barcelona, E-08193 Bellaterra, Spain
\and Universitat de Barcelona, ICC, IEEC-UB, E-08028 Barcelona, Spain
\and Japanese MAGIC Consortium, ICRR, The University of Tokyo, Department of Physics and Hakubi Center, Kyoto University, Tokai University, The University of Tokushima, KEK, Japan
\and Inst. for Nucl. Research and Nucl. Energy, BG-1784 Sofia, Bulgaria
\and Universit\`a di Pisa, and INFN Pisa, I-56126 Pisa, Italy
\and ICREA and Institute for Space Sciences (CSIC/IEEC), E-08193 Barcelona, Spain
\and now at Centro Brasileiro de Pesquisas F\'isicas (CBPF/MCTI), R. Dr. Xavier Sigaud, 150 - Urca, Rio de Janeiro - RJ, 22290-180, Brazil
\and now at NASA Goddard Space Flight Center, Greenbelt, MD 20771, USA and Department of Physics and Department of Astronomy, University of Maryland, College Park, MD 20742, USA
% ...
}
\date{Received;accepted}

\abstract
{The Geminga pulsar, one of the brighest gamma-ray sources, is a  promising candidate for emission of very-high-energy (VHE > 100 GeV) pulsed gamma rays. 
Also, detection of a large nebula have been claimed by water Cherenkov instruments. 
We performed deep observations of Geminga with the MAGIC telescopes, yielding 63 hours of good-quality data, 
and searched for emission from the pulsar and pulsar wind nebula. We did not find any significant detection, 
and derived 95\% confidence level upper limits. 
The resulting upper limits of $5.3 \times 10^{-13}$ TeV cm$^{-2}$ s$^{-1}$ for the Geminga pulsar and $3.5 \times 10^{-12}$ TeV cm$^{-2}$ s$^{-1}$ for the surrounding 
nebula at 50 GeV are the most constraining ones obtained so far at VHE. To complement the VHE observations, 
we also analyzed 5 years of Fermi-LAT data from Geminga, finding that the sub-exponential cut-off 
is preferred over the exponential cut-off that has been typically used in the literature. We also find that, 
above 10 GeV, the gamma-ray spectra from Geminga can be described with a power law with index softer than 5. 
The extrapolation of the power-law \emph{Fermi}-LAT pulsed spectra to VHE goes well below the MAGIC upper limits, 
indicating that the detection of pulsed emission from Geminga with the current generation of Cherenkov telescopes is very difficult.}


\maketitle

\section{Introduction}

Geminga is the first-known radio-quiet pulsar and the second
brightest persistent source in the GeV sky.
A review on the historical observations of Geminga can be found in \citealt{1996ARAA..34..331B}.
Its light curve exhibits two peaks, hereafter P1 and P2, separated by 0.5 in phase.
Gamma-ray emission from the interpulse region between P1 and P2 was reported by \citealt{fierro98}.
The period of Geminga ($P\sim 237$~ms) (\citealt{1992Natur.357..222H}) and its derivative
($\dot{P} \sim 1.1 \times 10^{-14}$~s/s) correspond to a spin-down age of
$\tau\sim 340$~kyr, a spin-down power $\dot{E}_{\rm rot} =
3.3\times10^{34}$~erg~s$^{-1}$ and a surface magnetic field $B_{\rm surf} \sim
1.6\times 10^{12}$~G. 
Although its spin-down luminosity is not as high as that of Crab and Vela, 
the short distance to this source makes the spin-down flux very large, 
which results in a high gamma-ray flux.

The mechanism of gamma-ray emission of pulsars is not yet fully understood.
Several emission locations were proposed as the origin of high-energy photons.
The polar cap (\citealt{1971ApJ...164..529S}, \citealt{1978ApJ...225..226H},  \citealt{1982ApJ...252..337D}) region,
located close to the neutron star surface in the open magnetosphere, was the first to be proposed.
However, a spectrum exhibiting a super-exponential cut-off at a few GeV is expected from the polar-cap gamma-ray emisson due to magnetic pair creation.
The proposed second region is the slot gap, located near the last open field line, which extends from the neutron star surface to
the null surface (\citealt{1983ApJ...266..215A}, \citealt{2003ApJ...598.1201D}, \citealt{2004ApJ...606.1143M}).
The third location of gamma-ray production proposed is the outer gap (\citealt{cheng1}, \citealt{cheng2}, \citealt{1995ApJ...438..314R}), which is 
located along the last open field line and extends from the null surface to the light cylinder.
The recent observations of the Crab pulsar at VHE by VERITAS and MAGIC (\citealt{2011Sci...334...69V}, \citealt{crabMagic25GeV}, \citealt{crab400GeV}, \citealt{crabBridgeMagic}) require a new model to explain the emission
above 100 GeV and the recent observation of pulsed emission above 400 GeV and extending beyond TeV energies, as reported recently by MAGIC (\citealt{2015arXiv151007048M}) challenges even more the theoretical models.
An extension of the outer-gap model has been proposed (\citealt{2012ApJ...754...33L}, \citealt{hirotani4}) in which the emission is explained by magnetospheric cascades inside the gap.
Recently, synchroton self-Compton emission from pairs was proposed to explain the emission from Crab, Vela and millisecond pulsars (\citealt{2015ApJ...811...63H}).
The pairs are created above the polar cap and absorb radio photons increasing their perpendicular momentum.
Another emission region, located at several light-cylinder radii, was investigated by \citealt{2000MNRAS.313..504B}, \citealt{aharonianWind},
where the pulsed X-ray photons are inverse Compton up-scattered by a cold ultra-relativistic pulsar-wind electrons.

Geminga was first detected as an unidentified gamma-ray source by the \emph{SAS-2} satellite (\citealt{SAS-2}).
In 1977, the \emph{COS B} satellite (\citealt{cosb1}) confirmed a gamma-ray emission from the same region.
In 1983 an X-ray counterpart of the \emph{COS B} source was observed (\citealt{1983ApJ...272L...9B}) and given the name \emph{Geminga}, and
in 1987 the optical counterpart was detected (\citealt{1987ApJ...319..358B}).
The X-ray pulsation was discovered by the ROSAT experiment (\citealt{1992Natur.357..222H}), 
and was further observed in gamma ray by the \emph{EGRET} telescope (\citealt{1992Natur.357..306B}) on board {\it Compton Gamma-Ray Observatory}, and \emph{COS B} (\citealt{1992Natur.357..287B}).
The first time-period derivative was estimated using {\it COS B} data (\citealt{1992Natur.357..287B}).
In 1981, the spectrum of Geminga was measured by the {\it COS B} satellite (\citealt{1981ICRC....1..177M}), being characterized by a simple power-law function from 100 MeV up to a few GeV.
The power-law spectrum was later confirmed by {\it EGRET} (\citealt{egretGeminga}) with a harder index.
The distance to the Geminga pulsar was first calculated by studying the interstellar absorption and proper motion, and was estimated to be approximately 100 pc (\citealt{1983ApJ...272L...9B}, \citealt{1993Natur.361..704B}).
Deeper study of the interstellar absorption, taking into account the spin-down properties of the pulsar set limit to the distance of Geminga to $250^{+150}_{-100}$ pc (\citealt{halpern2}).
Observations with the {\it Hubble Space Telescope} of the annual paralax led to more stringent constraint of the distance of $157^{+59}_{-30}$ pc (\citealt{1996ApJ...461L..91C}).\\

Event though the Geminga pulsar is radio quiet, several investigations were carried out to look for radio emission. 
A detection at 102.5 MHz was claimed in 1997 (\citealt{1997Natur.389..697M})
with a flux varying between 5 and 500 mJy.
Strong variations in the emission and pulse widths were reported too.
A soft spectrum would explain the absence of detection of pulsed emission above 102 MHz.
Recently, pulsed emission from the Geminga pulsar was reported at 42, 62 and 111 MHz (\citealt{malov}). 
From these recent observations, the previous radio silence from the Geminga pulsar has been interpreted as a long-term 
variability of the radio emission with a period of several years.\\

The Geminga pulsar, with one of the highest fluxes detected in the gamma-ray 
band (\citealt{3FGL}) ($4.5\times 10^{-9} \text{~erg cm}^{-2}\text{s}^{-1}$ above 100 MeV) and a spectrum extending above 25 GeV,
is a good candidate to be detected by Cherenkov telescopes. 
The detection of the Geminga pulsar with the MAGIC telescopes and the characterization of its timing and spectral 
features can shed light on the emission location and mechanisms at work in such an old pulsar.
 
One year of {\it Fermi}-LAT (Large Area telescope) (\citealt{2009ApJ...697.1071A}) 
observations at high energies resulted in a power-law spectrum with an exponential cut-off at ($2.5 \pm 0.2$) GeV (\citealt{Abdo10b}).
The study of the phase-resolved emission with fine binning shows a strong dependency of the cut-off energy
on the phase region considered. 
The pulsation is still clearly seen above 10 GeV with a reported significance greater than 6 $\sigma$, using 3 years of data,
and a hint was observed above 25 GeV (\citealt{catalogAbove10GeV}).

The spectral shape and the presence of the pulsed
emission above 25 GeV rules out the polar-cap model, in which a super-exponential cut-off is expected at a few GeV.
The \emph{Fermi}-LAT collaboration also reported that the peak intensity of P2 is getting stronger relative to the peak intensity of P1 above 200 MeV (\citealt{Abdo10b}).
Recently the VERITAS collaboration reported about the search for VHE 
emission from the Geminga pulsar with no signal detected above 100 GeV (\citealt{veritasGeminga}).
They computed upper limits of $4.0\times10^{-13} \text{cm}^{-2} \text{s}^{-1}$ and  $1.7\times10^{-13} \text{cm}^{-2} \text{s}^{-1}$
on the integrated flux above 135 GeV for P1 and P2, respectively, using a spectral index of $-3.8$.
The second catalog of hard \emph{Fermi}-LAT sources (2FHL) (\citealt{2015arXiv150804449T}), does not mention either the detection of Geminga above 50 GeV.

Besides the emission from the pulsar, an X-ray nebula was discovered around the Geminga pulsar (\citealt{caraveoNebula})
showing the presence of an extended structure aligned with the pulsar proper motion direction (\citealt{1993Natur.361..704B}).
Observations with the Chandra and XMM-Newton satellites (\citealt{Luca06,2006ApJ...643.1146P}) reported the 
detection of three tail-like structures behind the pulsar; one 25'' tail 
aligned to the pulsar proper motion, and two 2' outer tails. Another 50'' emitting region ahead of the pulsar was reported.

At gamma-ray energies, the {\it Fermi}-LAT reported a continuous emission over the whole pulsar rotation, but it is incompatible with a surrounding nebula (\citealt{Abdo10b}).
The Whipple collaboration obtained an integral flux upper limit for
continuous emission of $8.8\times10^{-12}$~cm$^{-2}$~s$^{-1}$ above 0.5~TeV (\citealt{1993AA...274L..17A}). %
At higher energies, the Milagro collaboration reported the detection of a TeV
extended steady emission from Geminga at a significance of
6.3 $\sigma$, recently confirmed by HAWC (\citealt{gemingaHAWK}).
Milagro observed an emission region that is extended by 2-3
degrees and reported a flux level of $(38\pm 11)\times10^{-17}$~TeV$^{-1}$~cm$^{-2}$~s$^{-1}$
at 35~TeV (\citealt{Abdo09}). 
At radio frequencies, many observers have attempted to detect a continuous emission from Geminga.
Only the deepest VLA interferometric observation of Geminga performed in 2004 (\citealt{vlaGeminga}),
resulted in the detection of steady radio emission.
Overall, the Geminga radio tail is compatible with the scenario of a synchrotron-emitting PWN. \\
~\\
In order to study the gamma-ray emission of the Geminga pulsar and nebula, we collected 75 hours of observation with MAGIC. 
Furthermore, we performed the analysis of 5 years of \emph{Fermi}-LAT data in order to complement the VHE observations.

%
%
\section{MAGIC observations and data analysis}
The MAGIC telescopes are a set of two imaging atmospheric Cherenkov telescopes.
They are located  at a height
of 2200 m a.s.l. in the Roque de los Muchachos Observatory, on
La Palma island (Spain).
Both telescopes consist of a 17 m diameter reflector
and a fast imaging camera with a field of view of $3.5^{\circ}$ diameter. 
The trigger threshold for standard observations at zenith angles below $35^{\circ}$ is 
around 50 GeV.
The MAGIC telescopes have an integral sensitivity of $0.66\%$ of the Crab Nebula flux above 220 GeV for 50 hours of observation, 
with an angular resolution of $\sim 0.07^{\circ}$ and an energy resolution of 16\% (\citealt{performance}).

Observations of the Geminga pulsar and nebula were performed
between December 2012 and March 2013, with the upgraded MAGIC telescopes (\citealt{magicUpgrade}).
During this period, a total of  $\sim$ 75 hours were taken 
at zenith angles below $35^{\circ}$ to ensure the lowest possible energy threshold.
The observations were performed in the so-called wobble mode (\citealt{1994APh.....2..137F}), 
where the source is offset by $0.4^\circ$ from the camera center. 
After rejection of data taken under unfavorable weather or technical conditions, 63 hours
of data remained for the analysis.
Together with each event image, we recorded the absolute event arrival time 
using a GPS receiver. The performance of the MAGIC
time acquisition system was evaluated by observing periodically the 
Crab pulsar in the optical wave band with a special PMT
located at the MAGIC camera center (\citealt{cpix}).
~\\
The data analysis was performed using the standard MAGIC analysis chain {\it MARS} (\citealt{ZaninMars}).
The phase of the events was computed using {\it tempo2} (\citealt{tempo2}).
The ephemeris was provided by the {\it Fermi}-LAT collaboration\footnote{\small http://www.slac.stanford.edu/$\sim$kerrm/fermi\\\_pulsar\_timing/J0633+1746/html/J0633+1746\_54683\_56587\_chol.par} (\citealt{2011ApJS..194...17R}).
For the pulsar analysis, gamma-ray candidate events are selected by applying cuts in hadronness and in $\theta^{2}$. 
Hadroness is a particle-identification estimator that classifies events into gamma-ray or hadron candidates, 
while $\theta^{2}$ is the squared angular distance between the source position and the re-constructed source position.
The cuts are optimized using a background sample and Monte Carlo 
gamma-ray sample by maximizing in each energy bin the Q-factor defined as: $Q=\epsilon_{on}/\sqrt{\epsilon_{off}}$, 
where $\epsilon_{on}$ and  $\epsilon_{off}$ are the efficiency of the cuts for signal and background data, respectively.
For the computation of the cuts we imposed that at least 50\% of the Monte Carlo gamma-ray events survive the cuts.
The significance of the pulsed emission was estimated using equation 17 in \cite{LiMa}.
The upper limits on the pulsed emission were computed using 
the Rolke method (\citealt{rolke1}) assuming a Poissonian background and requiring a 95\% confidence level.

The search for a steady extended emission was done computing the signal to 
noise ratio around the Geminga pulsar. Several extensions around the Geminga pulsar were considered, 
setting different value of the cut in  $\theta^{2}$  (0.04, 0.06, 0.08 and 0.1 deg$^{2}$).
Also, a significance sky map of the region around the Geminga pulsar was produced.
The significance in each bin of the sky map was computed using the Li~\&~Ma method applied
on a background estimate.
The cuts were selected maximizing the Q-factor on a contemporaneous Crab Nebula sample
using the hadronness and size parameters of the images, which is defined as the sum of the charge from each pixel.
The upper limits for the nebula emission were computed using the same method as for the pulsed emission and different spectral assumption.

%
%
\section{\emph{Fermi}-LAT observation}
\subsection{Fermi data analysis}
A data sample of 5 years 
(from 54710 up to 56587 MJD) of {\it Fermi}-LAT data was analyzed.
We analyzed this data-set using the P7REP\_SOURCE\_V15 instrument response functions and the {\it Fermi} tools version v9r31p1.
We selected events that were recorded when the telescope was in nominal science mode and when the rocking angle was 
lower than $52^{\circ}$.
To reject the background coming from the Earth's limb, we selected photons with a zenith angle $ \leq 100^{\circ}$.
The phase and barycentric corrections of the events were computed using {\it tempo2} 
using the same ephemeris as for MAGIC data.
We computed the light curve and the spectral energy distribution for both peaks, P1 and P2, separately. 
Furthermore, we calculated the phase averaged (PA) emission.
The pulsar light curve was produced using an energy dependent region of interest (ROI) with a radius defined as 
$R = \max(6.68 - 1.76\times \log(E),1.3)^{\circ}$   as done in \citealt{2010ApJ...708.1254A}.

For the spectral analysis, the binned likelihood method was used. We set the ROI to $15^{\circ}$ as done in \citealt{2pulsarCatalog}. 
We included all the sources from the third {\it Fermi} catalog (\citealt{3FGL}) in the background model. 
For sources with a significance  higher than 5 $\sigma$ and located at less than 10 degrees away from the 
Geminga pulsar, only  the normalization factor was left free.
We also let the normalization factor of the isotropic and Galactic background models free.
We discarded all the sources with TS < 2.
For all the remaining sources all the parameters were fixed to the catalog values.
For the calculation of the spectral points, we repeated the procedure in each energy bin 
using a power law with the spectral index and normalization factor free. 
Only spectral points with a significance higher than 2 $\sigma$ are shown on the plots.

\subsection{Fermi-LAT results}
\begin{figure}[h!]
\hspace{-0.5cm}
\centering
  \includegraphics[scale=0.5]{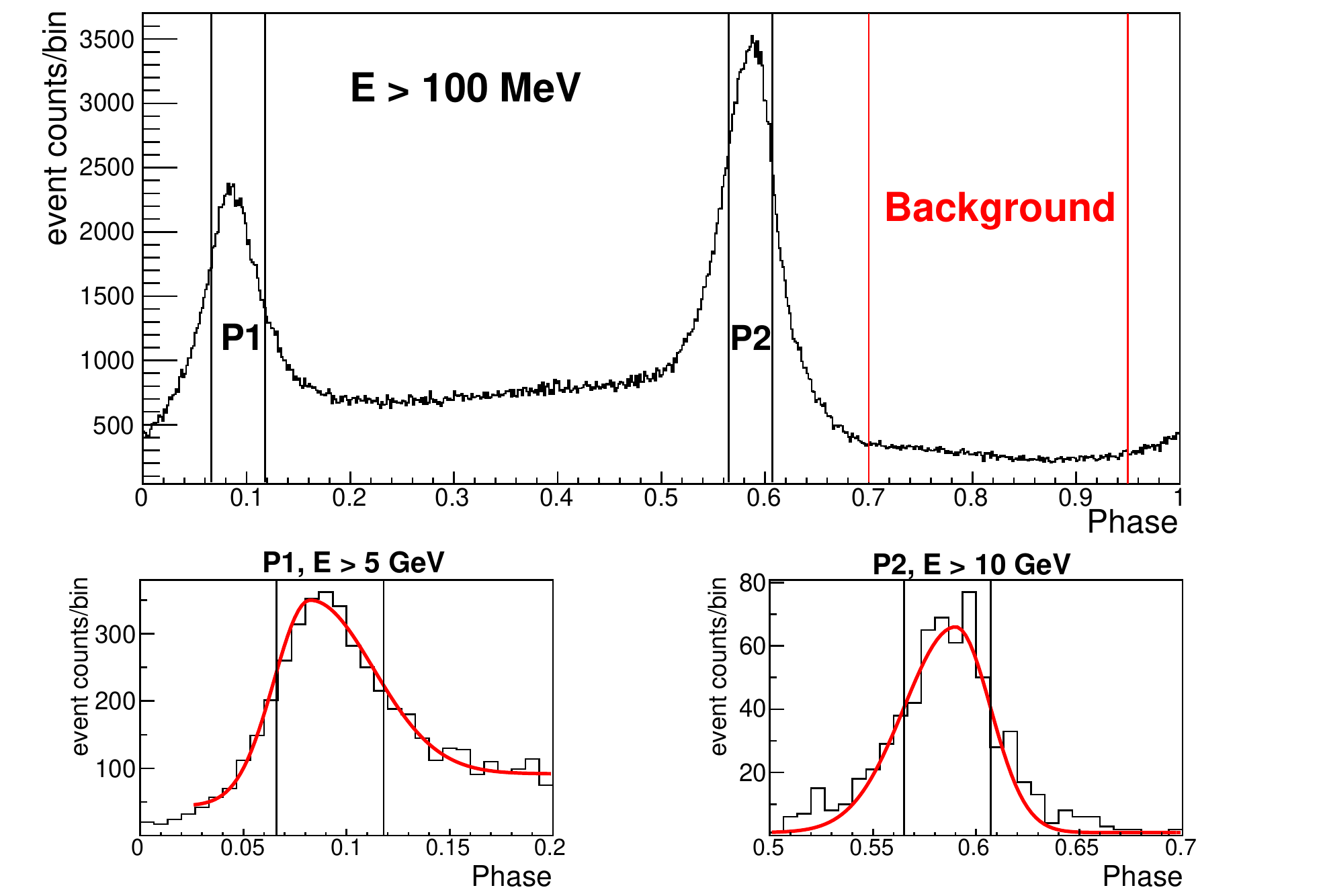}
  \caption{\small Light curve computed with the {\it Fermi}-LAT data above 100 MeV (top). 
    A close-up is made on both P1 above 5 GeV and P2 above 10 GeV and their corresponding gaussian fits (bottom), with resulting $\chi^{2}$/d.o.f values of 61/26 
    and 32/29 for P1 and P2, respectively. The vertical black lines define the signal regions, while the vertical red lines define the off-pulse region used to determine the background}
   \label{fitPeaks}
 \end{figure}

We computed the light curve above 100 MeV.
To determine the pulses profiles and OFF phase range we used photons with energy larger than 5 GeV for P1 and 
larger than 10 GeV for P2. 
The two different energy ranges are motivated by the aim of evaluating the peak shape at the largest energy, for a better matching with the one
we would expect at the MAGIC energy range, maintaining enough statistics.
We fit both peaks to asymmetric Gaussian functions. 
We used as signal region the peak position $\pm 1\sigma$, as shown in Table \ref{fermiFit}.
We defined the background region in the off-phase where no emission is expected from the pulsar.
From now on P1 and P2 will always be referred as the values in Table \ref{fermiFit}.
The obtained light curve above 100 MeV together with the signal and background regions are shown in Figure \ref{fitPeaks}
together with a close-up on the fits of P1 and P2 at the corresponding energies.\\
\begin{table}[ht]
 \centering 
 \caption{\small Definition of the signal and off-pulse regions derived from the LAT data.}
\begin{tabular}{ccc}
 \hline 
P1& P2 & Off-region\\
\hline
\hline
0.066 - 0.118 & 0.565 - 0.607 & 0.7 - 0.95\\

 \hline
 \end{tabular}
\label{fermiFit}
 \end{table}

We fit the spectral energy distribution (SED) of P1, P2 and the phase-averaged (PA) using two different spectral shapes: power-law with an exponential cut-off function (EC), 
and power-law with a sub-exponential cut-off function (SEC).
The sub-exponential ($b$<1)  and exponential ($b$=1) cut-off functions are defined by the following equation:
\begin{eqnarray}
\frac{dF}{dE} = N_{0}\left(\frac{E}{E_{0}}\right)^{-\alpha}\exp(-(E/E_{c})^{b}),
\end{eqnarray}
where $E_{0}$ is the energy scale, set to $927.9$ MeV as computed in \citealt{3FGL}, $\alpha$ the spectral index, and $E_{c}$ the cut-off energy.
The results of the computed spectra using a SEC function are shown in Table \ref{fermiSpectra}.
\begin{table}[h!]
\centering
\caption{\small  Spectral parameters of the fit using the likelihood method for the SEC 
function between 100 MeV and 100 GeV for P1, P2 and PA.
The normalization factors, $N_{0}$, are given in unit of $10^{-10}$MeV$^{-1}$·s$^{-1}$·cm$^{-2}$. The quoted errors are statistical at a 1 $\sigma$ confidence level. The systematic errors reported by the \emph{Fermi}-LAT are
of 14\% on $\alpha$ and 4\% on $E_{c}$ (\citealt{2pulsarCatalog}).}

\begin{tabular}{ccccc}
\hline
~  &   $N_{0}$  &  $\alpha$ &  $E_{c}$ [GeV] & $b$ \\
\hline
\hline
 P1 & $3.0 \pm 0.3 $  & $1.12 \pm 0.04$ & $1.2 \pm 0.1$  & $0.81 \pm 0.04$  \\ 
 P2 & $4.3 \pm 0.4 $  & $0.78 \pm 0.03$ & $1.1 \pm 0.1$  & $0.70 \pm 0.03$  \\ 
 PA & $28.3 \pm 1.8 $ & $0.94 \pm 0.02$ & $0.8 \pm 0.1$  & $0.67 \pm 0.02$  \\ 

\hline
\end{tabular}
\label{fermiSpectra}
\end{table}

In order to characterize the emission at high energies, 
we fit the high-energy tail (above 10 GeV) for both P1 and P2 using a power law.
The normalization factors were computed at 10 GeV.
The results of the power-law fit above 10 GeV are shown in Table \ref{table::fermiFit}.

\begin{figure}[ht]
  \centering
  \includegraphics[scale=0.45]{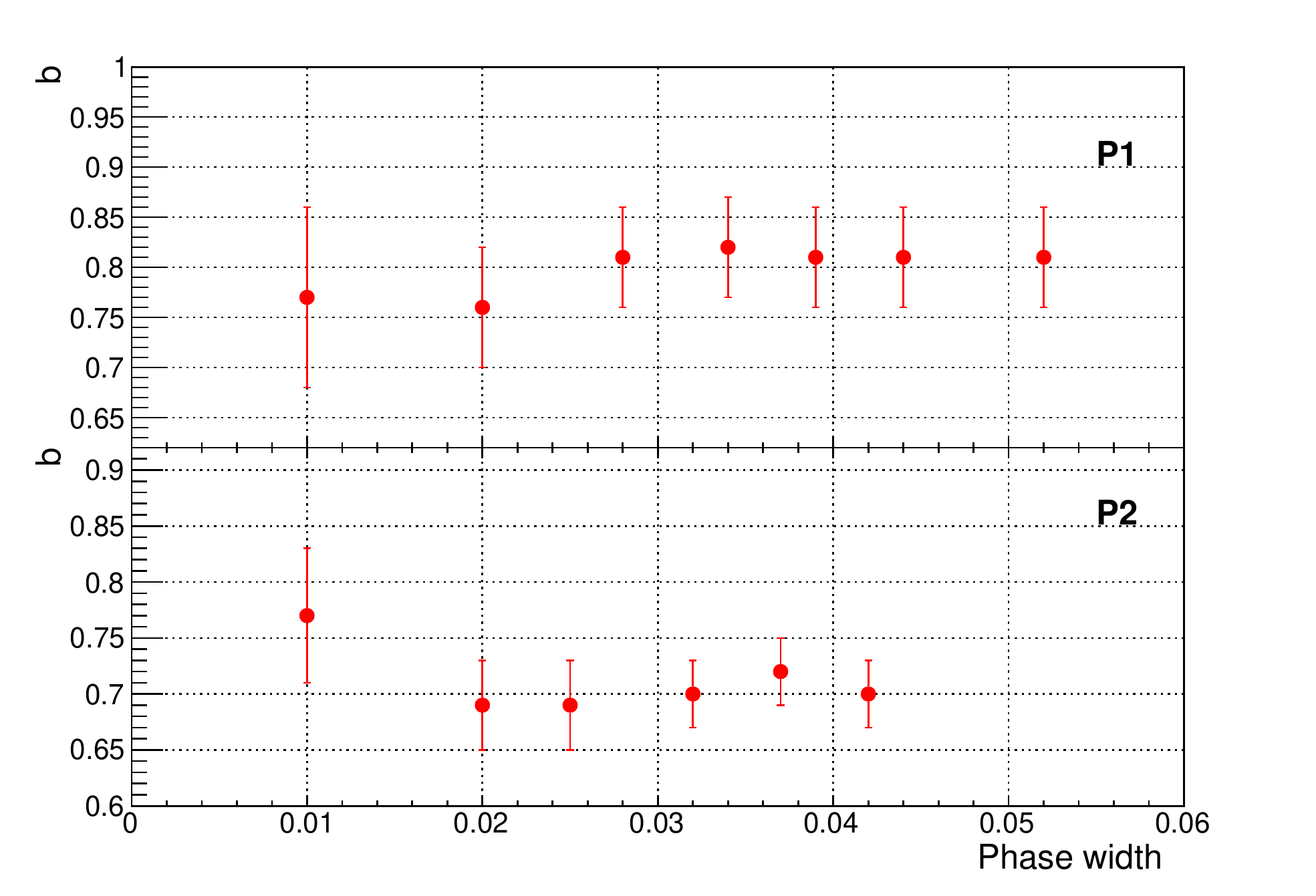}
  \caption{\small Computation of the sub-exponential cut-off  $b$ parameter according to the phase 
width considered for P1 (top) and P2 (bottom).}
   \label{figure::b-peakWidth}
 \end{figure}

\begin{table}[h!]
\begin{center}
\caption{\small Results of the fit of P1 and P2 spectral energy distribution with a power law above 10 GeV.
The normalization factor, $N_{0}$, is given in unit of $10^{-9}$MeV$^{-1}$·s$^{-1}$·cm$^{-2}$. }
\begin{tabular}{ccc}
\hline
~ &$N_{0}$ & $\alpha$  \\
\hline
\hline
P1  & $(5.9 \pm 1.4)\times 10^{-5}$ & $5.3 \pm 0.7$ \\

P2  & $ (7.2 \pm 0.1)\times 10^{-4}$ & $5.2 \pm 0.3$ \\
\hline
\end{tabular}
\label{table::fermiFit}
\end{center}
\end{table}
The resulting spectra computed using 5 years of {\it Fermi}-LAT data are consistent with 
the previous results reported by the {\it Fermi}-LAT collaboration (\citealt{Abdo10b}, \citealt{2pulsarCatalog}).
The SEC appears to be in better agreement with the data. 
The $b$ parameter, indicating how much from an EC the data deviates, is significantly smaller than one. 
Also, the calculation of the likelihood ratio of the SEC model over EC results in a deviation for the SEC of 6 $\sigma$, 11 $\sigma$ and 24 $\sigma$ for P1, P2 and PA, respectively.
We also computed the SED using finer binning in order to estimate the evolution of the $b$ parameter according
to the phase width considered.
The top and bottom panels in Figure \ref{figure::b-peakWidth} represent the value computed for P1 and P2, respectively.
The smallest width was taken as 0.01 in phase due to the lack of statistics for smaller regions.
We did not observe any significant variation of $b$ with the pulse width.

%
%

\section{MAGIC Results}
\begin{figure}[h!]
  \centering
  \includegraphics[scale=0.45]{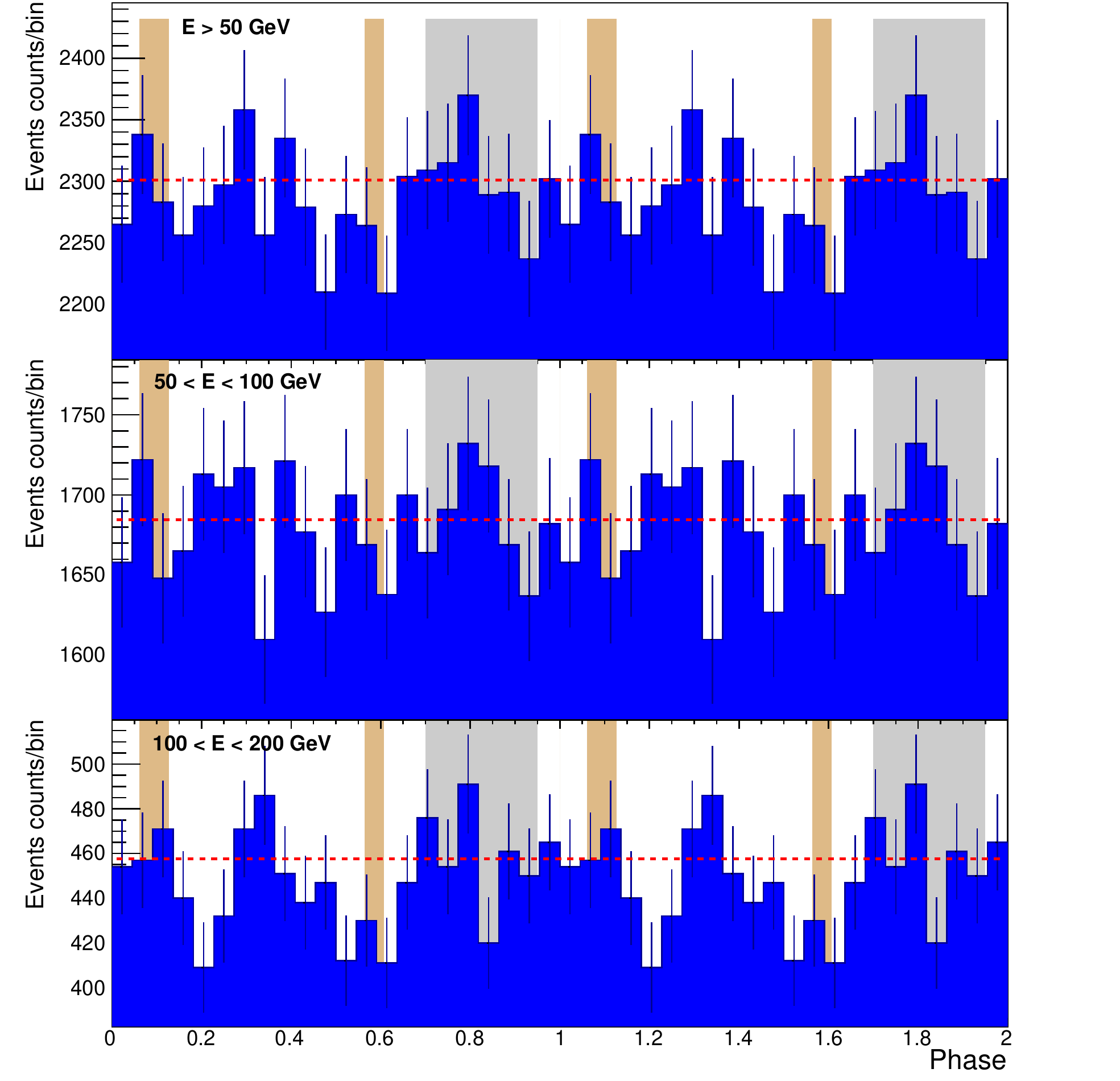}
  \caption{\small Light curves of the Geminga pulsar obtained with MAGIC for different energy bins. From top to bottom: above 50 GeV, 50-100GeV and 100-200 GeV. 
    Two cycles are plotted for clarity.
    The bin width corresponds to $\sim$ 10.8 ms (1/22 of the Geminga rotational period).
    The shaded brown areas show the positions of P1 (main pulse) and P2
    (interpulse). The grey area shows the off-region. The dashed red line represents the averaged number of events in the background region.}
   \label{GemingaPulsed}
 \end{figure}

We computed the light curve and the corresponding significances for the pulsed emission in three energy ranges: above 50 GeV, 50--100 GeV and 100--200 GeV,
as shown in Figure \ref{GemingaPulsed}.
The background is estimated from the off-pulse region (grey area; phase 0.70 -- 0.95) and
the dashed red line represents the averaged number of events in the background region
We computed the significance for P1, P2, and the sum of both peaks.
The results of the statistical tests are shown in Table \ref{table_significances}.
No significant pulsation was found in MAGIC data in any of the energy ranges investigated.
We computed the upper limits for the pulsed emission. 
The spectral indices used for the upper limits computation were obtained from the extrapolation of P1 and P2 \emph{Fermi}-LAT spectra above 10 GeV using a powe law (see Table \ref{table::fermiFit}).

\begin{figure}[h!]
    \includegraphics[scale=0.43]{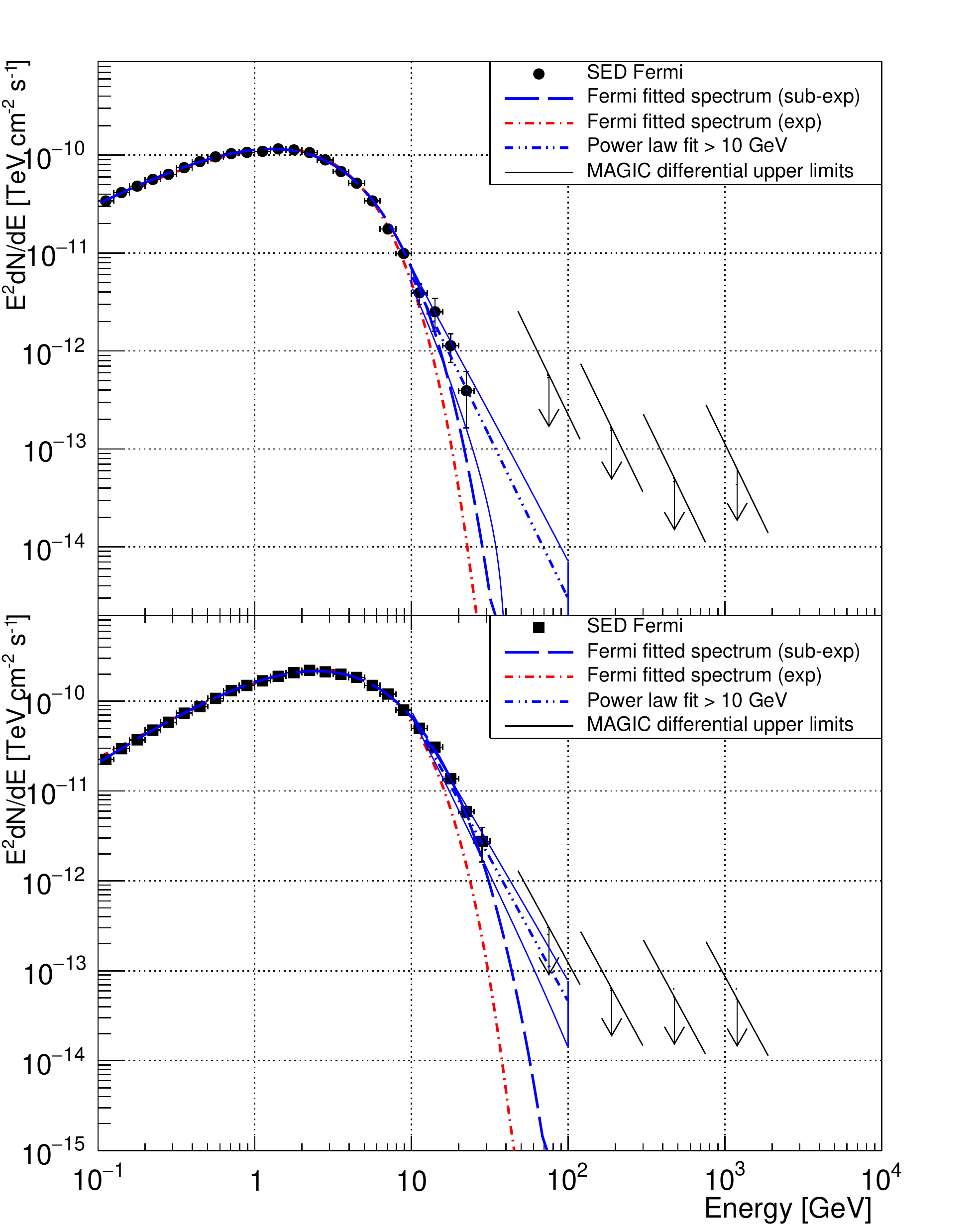}\\
  \caption{\small P1 (top) and P2 (bottom) SED.
  The differential upper limits are represented by the black arrows. The blue dashed line represents the SED computed using 5 years of {\it Fermi}-LAT data assuming a SEC function, between 100 MeV and 100 GeV, 
and the dot-dashed line the fit of SED to a EC function.
  The dot-dot-dashed line is the result of the fit of the Fermi data above 10 GeV with a power law. The statistical error contour from the power-law fit is also plotted.}
\label{peaksSED}
 \end{figure}

The differential upper limits computed for the pulsed emission are shown in Figure \ref{peaksSED} by the black arrows.
The black lines on top of the arrows have the spectral slope 
used for the upper limit computations.
The dot-dot-dashed blue line represents the fit to {\it Fermi} data above 10 GeV with a power-law function
, the dashed line the results of the fit of the SED to SEC and the dot-dashed line the result of the fit of the SED to EC.
The statistical error contour is also plotted for 
the power-law fits at high energies.

\begin{table}[ht]
\centering
\caption{\small Significance computed for P1, P2 and the sum of both peaks.
The significances were computed using Li~\&~Ma.}
\begin{tabular}{cccc}
\hline
Energy range (GeV)  & P1 & P2 & P1 + P2 \\
\hline
\hline
$\geq 50$ & $0.2 \sigma$  & $-0.1 \sigma$ & $0.1 \sigma$  \\ 
50-100    & $-0.2 \sigma$ & $0.2 \sigma$ & $0.0 \sigma$   \\ 
100-200   & $0.7 \sigma$  & $-1.4 \sigma$ & $-0.3 \sigma$ \\ 
\hline
\end{tabular}
\label{table_significances}
\end{table}

\begin{figure}[h!]
   \centering
  \includegraphics[width=0.37\textwidth]{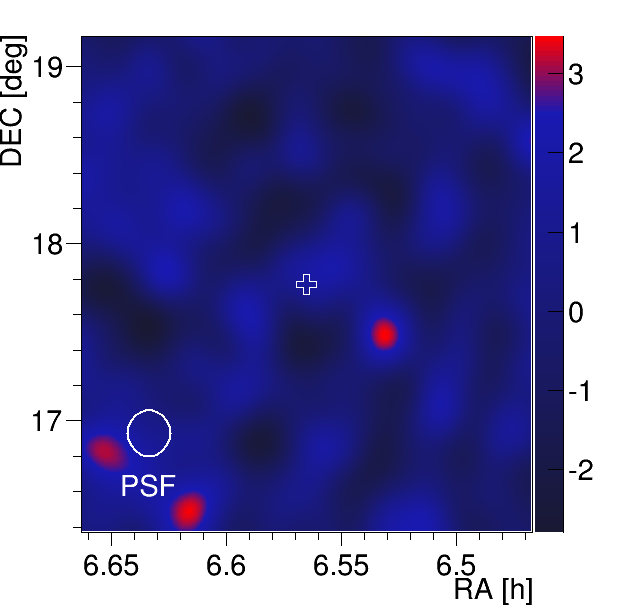}
  \caption{\small Sky map representing the signal significance computed around the location of the Geminga pulsar using MAGIC data above 50 GeV. 
    The cross at the center of the map represents the Geminga pulsar location. The white circle represents the function used for the deconvolution of the sky map.}
   \label{skymaps}
 \end{figure}

\begin{figure}[h!]
  \includegraphics[scale=0.45]{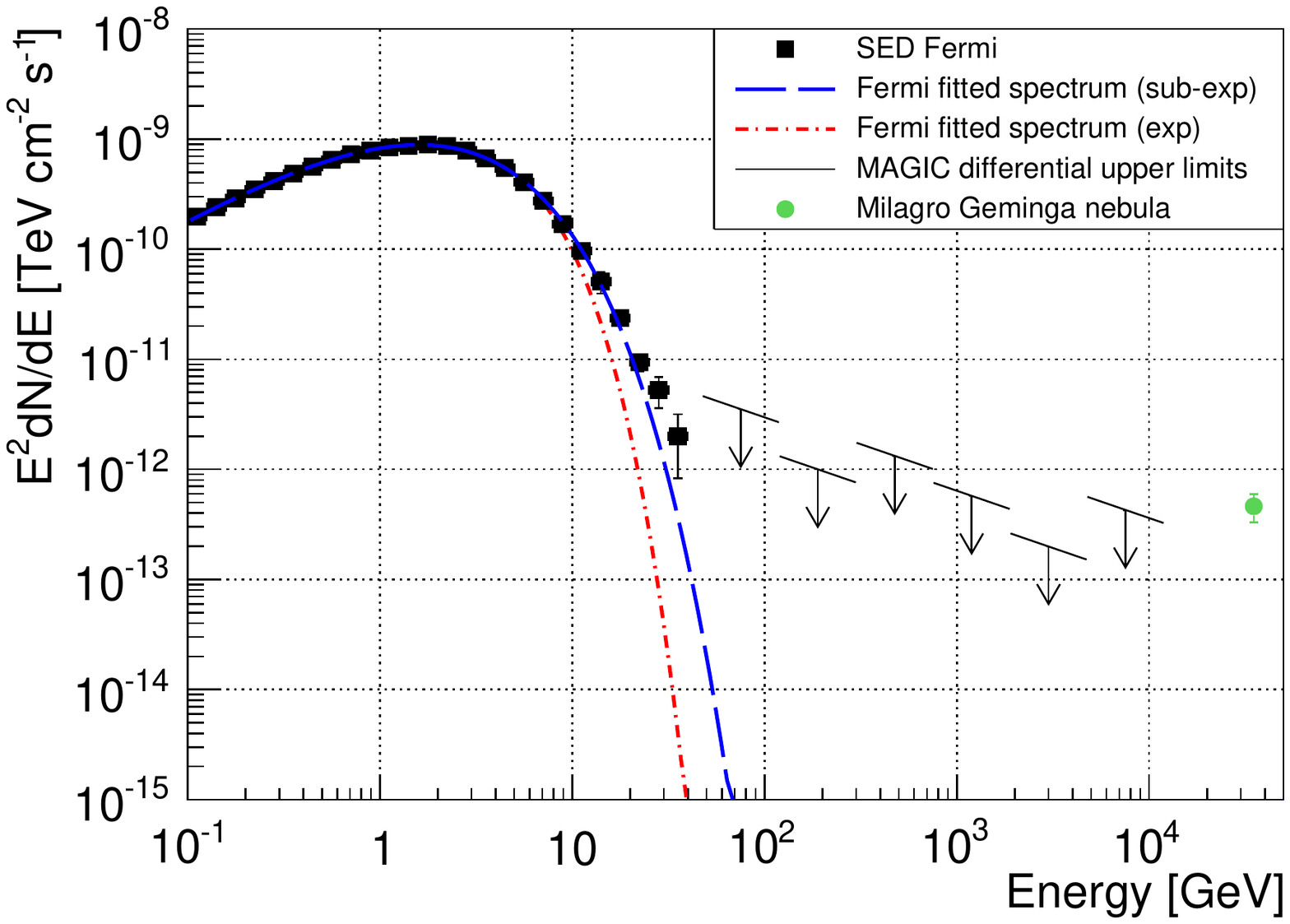}
  \caption{\small Phase averaged spectral energy distribution.
    The differential upper limits are represented by the black arrows. The blue dashed lines represents the SED computed using 5 years of {\it Fermi}-LAT data assuming a SEC function, between 100 MeV and 100 GeV.
    The green point represents the flux level of the Geminga Nebula as seen by MILAGRO.}
  \label{phaseAveragedSED}
\end{figure}

Figure \ref{skymaps} shows the sky map of the signal significance around the Geminga
pulsar for the steady emission using MAGIC data.
The position of the Geminga pulsar is marked with a cross.
The white circle represents the standard deviation of the Gaussian function used for the smearing of the sky map.
No significant emission was found from the Geminga nebula above 50 GeV.
~\newline
%
%
We calculated the differential upper limits on the emission 
from the nebula surrounding the Geminga pulsar
in the energy range covered by MAGIC.
The computed differential upper limits are represented by the black arrows 
in Figure \ref{phaseAveragedSED}.
The spectral index used for the upper limit computation was taken as $-2.6$.
In order to estimate the  upper limit variations due to the assumption of the spectral index value, 
we recomputed the upper limits assuming two different spectral indices of $-2.0$ and $-2.8$.
The two chosen values define the typical range of spectral indexes for pulsar wind 
nebulae (\citealt{kargaltsevBook}).
A fluctuation of 13\% is observed in the upper limit computation below 120 GeV. 
For energies above 120 GeV the variations are below 10\%.
We also estimated the integral upper limits on the emission from the nebula to be $2.4\times 10^{-11}$cm$^{-2}$s$^{-1}$ and $3.2\times 10^{-12}$cm$^{-2}$s$^{-1}$
above 50 GeV and 200 GeV, respectively.
In Figure \ref{phaseAveragedSED}, the computed PA SED using 5 years of {\it Fermi}-LAT data is represented by the black points. 
The dashed blue lines is the result of {\it Fermi} spectral shape computation using a power-law 
with a sub-exponential cut-off
and the dot-dashed line using a power-law with an exponential cut-off.
The green point represents the flux level of the Geminga Nebula as seen by MILAGRO (\citealt{Abdo09}).
%
%
\section{Discussion and conclusions}
During the winter 2012/13, the Geminga pulsar and its surrounding nebula were observed for 63 good-quality-selected hours by
the MAGIC telescopes to search for emission from the pulsar and its surrounding 
nebula at VHE. The analysis of the MAGIC data yielded no significant signal and hence 
resulted the computation of upper limits above 50 GeV for both pulsed and 
steady emission. 
Besides MAGIC data, 5 years of \emph{Fermi}-LAT data were analyzed to derive pulsed and phase-averaged emission and compare to the VHE upper limits. 
Our results on the analysis of MAGIC and \emph{Fermi}-LAT data are consistent with those reported in the 2FHL, where no significant signal from Geminga was found above 50 GeV.
In addition, the computed integral upper limit on the emission from the nebula above 200 GeV is compatible with the flux level reported by Milagro.

The \emph{Fermi}-LAT spectra from 0.1 GeV to 30 GeV can be described by a power-law with a sub-exponential cut-off. As reported by (\citealt{lyutikov}), 
a simple power-law can also be used to characterize the emission at the high energies, and more statistics would be 
required to distinguish between the spectral shapes. The upper limits computed using the MAGIC data are well above the 
\emph{Fermi}-LAT power-law spectra extrapolated to VHE, and hence they do not provide additional constrains to the spectral shape of the pulsed emission. 
Therefore, the mechanism responsible for the high-energy emission from the Geminga pulsar is difficult to establish. 
At high energies, the emission due to synchro-curvature radiation and inverse Compton scattering are expected to 
exhibit different spectral shapes.
For example, in the framework of the outer-gap model, where the high-energy emission takes place at high altitudes from the neutron 
star (\citealt{cheng1}, \citealt{cheng2}), a curvature or synchro-curvature radiation mechanism would exhibit a spectral shape well characterized 
by  an exponential cut-off (\citealt{curvatureAharonian}, \citealt{2015MNRAS.449.3755V}). As the radiation is very sensitive to 
the pitch angle of the radiating particles, the sum of the emission from particles with the same energy but different angles results in a less abrupt cut-off.
Furthermore, calculations of the outer-gap magnetic-field-aligned electric field evolution (\citealt{hirotani2}, \citealt{hirotani4}) 
show that the accelerating electric field depends on the height in the gap and reaches a maximum
in the center of the gap. Distinct heights with different values of the electric field would accelerate particles at different energies, 
resulting in a spread of the cut-off energy values.
A strong dependency of the cut-off energy on the accelerating electric field is reported by (\citealt{syncroCurveDiego}).
Such a behavior of the cut-off values was reported for the Geminga pulsar (\citealt{Abdo10b}).
The {\it Fermi} collaboration studied the phase-resolved evolution of the cut-off energy for the Geminga pulsar over the whole pulsar rotation
using bin sizes such as each bin contains 2000 photons.
The results show that within the P1 and P2 phase regions, where the computed cut-off values are the highest, these values vary.
Considering wider phase ranges, the fluctuations
of the cut-off value would result in an a sub-exponential cut-off spectral shape.
However, the pulsed gamma-ray spectra we computed using fine bins in phase around the pulses’ positions discard the 
exponential cut-off because the best fit values for the $b$ parameter are significantly smaller than 1.
In the case of synchro-curvature radiation, this deviation can arise from the caustic emission (\citealt{2003ApJ...598.1201D}), i.e, 
overlapping of photons emitted at different heights and along different magnetic field lines.
The caustic effect being more important for P2 than P1, due to the curvature of the magnetic field line, would 
explain the greater values of $b$ for P1 with respect to P2.

In the case of an inverse Compton (IC) emission or synchroton self-Compton within the outer gap (\citealt{hirotani4}), the break in the spectral shape would correspond
to a break in the particle distribution function (\citealt{lyutikov}) if all the emission comes from this mechanism.
If the particles are distributed as a broken power law, then the IC spectrum would appear as a broken power law too, 
and a high-energy power-law like tail would be seen as it is the case for the Crab pulsar (\citealt{crabVeritas}, \citealt{crabMagic25GeV}, \citealt{crab400GeV}, \citealt{crabBridgeMagic}, \citealt{2015arXiv151007048M}).
However, in the case of an inverse Compton emission, the power-law tail exhibited by the Geminga pulsar would be much
softer than that of the Crab (\citealt{crabBridgeMagic}), as can be seen from the power-law spectral fit of the {\it Fermi}-LAT
data above 10 GeV. 
A hard gamma-ray tail is not expected even if the curvature radiation is produced in a curved magnetic field close to the light cylinder (\citealt{2012MNRAS.424.2079B}).\\
The analysis of the nebula around the Geminga pulsar shows no significant detection above 50 GeV.
The presence of the nebula is unknown at the GeV scale.  Indeed, the observations of the Geminga pulsar with the
{\it Fermi}-LAT show no evidence of a surrounding nebula. 
The detection of a large nebula similar to the one claimed by the Milagro Collaboration is not straightforward for MAGIC, 
as its extension is larger than the field of view of the telescopes.
Overall, the prospects of detecting the Geminga pulsar with the current Cherenkov telescopes are rather low. 
However, the upcoming Cherenkov Telescope Array (CTA)(\citealt{cta}) could, with a better sensitivity and a lower energy threshold, 
detect high energy gamma-ray emission from the Geminga pulsar and thus shed light on the physics of pulsars. 
We have estimated that Geminga could be detected at a 5 $\sigma$ level by CTA in 50 hours.

\section{Acknowledgments}

We would like to thank
the Instituto de Astrof\'{\i}sica de Canarias
for the excellent working conditions
at the Observatorio del Roque de los Muchachos in La Palma.
The financial support of the German BMBF and MPG,
the Italian INFN and INAF,
the Swiss National Fund SNF,
the ERDF under the Spanish MINECO (FPA2012-39502), and
the Japanese JSPS and MEXT
is gratefully acknowledged.
This work was also supported
by the Centro de Excelencia Severo Ochoa SEV-2012-0234, CPAN CSD2007-00042, and MultiDark CSD2009-00064 projects of the Spanish Consolider-Ingenio 2010 programme,
by grant 268740 of the Academy of Finland,
by the Croatian Science Foundation (HrZZ) Project 09/176 and the University of Rijeka Project 13.12.1.3.02,
by the DFG Collaborative Research Centers SFB823/C4 and SFB876/C3,
and by the Polish MNiSzW grant 745/N-HESS-MAGIC/2010/0.


\bibliography{biblio}
\bibliographystyle{aa}

\end{document}